\documentclass[12pt]{iopart}
\pdfoutput=1
\usepackage{epsfig}
\usepackage{graphics}
\usepackage[usenames]{color}
\usepackage{verbatim}
\usepackage{graphicx}
\usepackage{amssymb,iopams,cite}

\newcommand{\beq}{\begin{equation}}
\newcommand{\eeq}{\end{equation}}
\newcommand{\bx}{{\bf x}}
\newcommand{\ii}{{\rm i}}
\newcommand{\ee}{{\rm e}}

\newcommand{\revision}[1]{#1}

\begin{document}

\title{Glass to superfluid transition in dirty bosons on a lattice}
\date{\today}
\author{Julia Stasi\'nska$^{1,\#}$, Pietro Massignan$^{2}$, Michael
Bishop$^{3}$,\\ Jan Wehr$^{3}$, Anna Sanpera$^{1,4}$ and Maciej
Lewenstein$^{2,4}$}
\address{$^1$ F\'isica Teorica: Informaci\'o i Processos Qu\`antics,
Universitat Aut\`onoma de Barcelona, 08193 Bellaterra (Barcelona), Spain.}
\address{$^2$ ICFO-Institut de Ci\`encies Fot\`oniques, Mediterranean Technology
Park, 08860 Castelldefels (Barcelona), Spain.}
\address{$^3$ Department of Mathematics, University of Arizona, Tucson, AZ
85721-0089, USA.}
\address{$^4$ ICREA - Instituci\'o Catalana de Recerca i Estudis Avan\c{c}ats,
08010 Barcelona, Spain.}
\ead{$^{\#}$ julsta@ifae.es}

\begin{abstract}
We investigate the interplay between disorder and interactions in
a Bose gas on a lattice in presence of randomly localized
impurities. We compare the performance of two theoretical methods,
the simple version of multi-orbital Hartree--Fock and the common
Gross--Pitaevskii approach, showing how the former gives a \revision{very good}
approximation to the ground state in the limit of weak
interactions, where the superfluid fraction is small. We further
prove rigorously that for this class of disorder the fractal
dimension of the ground state $d^*$ tends to the physical
dimension in the thermodynamic limit. This allows us to introduce
a quantity, the fractional occupation, which gives insightful
information on the crossover from a Lifshits to a Bose glass.
Finally, we compare temperature and interaction effects,
highlighting similarities and intrinsic differences.
\end{abstract}


\submitto{\NJP}
\maketitle

%%%%%%%%%%%%%%%%%%%%%%%%%%%%%%%%
\section{Introduction}
The physics of disordered quantum systems is a very active field
of research since the 1950s, when this topic received the
fundamental contributions of Anderson and Mott~\cite{Kramer93}.
Their works showed that, in presence of disorder, even ideal
conductors undergo a phase transition towards an insulating state
due to destructive quantum interference. An ideal playground for
the investigation of quantum effects is offered by ultracold
atoms, where a controllable amount of disorder may be implemented
in many ways, e.g., by means of speckle potentials, bichromatic
lattices with incommensurate frequencies, localized impurities, or
site-resolved addressing in optical lattices. After a long search,
Anderson localization of ultracold ideal gases was finally
observed in one dimension (1D) \cite{Anderson1Dexp}, and very
recently reported also in 3D \cite{Anderson3Dexp}.

At this point it is important to stress the fundamental difference
between disordered fermionic and bosonic systems.
For fermions the crucial role is played by the Fermi energy: the
question whether a given system is a conductor or insulator
depends on whether the single particle states close to Fermi level
are localized or not. Bosons, in contrast, tend to condense in the
lowest energy state (or states). The question whether they
constitute a superfluid or an insulator reduces then to
understanding if the low-energy states are localized or not (for
more extensive discussion see for instance
\cite{Sanchez-Palencia10,Lewenstein12}).

%%%%%%%%%%%%%%%%%%%%%%%%%%%%%%%%%%%%%%%%%%%%%%%%
\begin{figure}[!ht]
\centering
\includegraphics[width=0.5\textwidth]{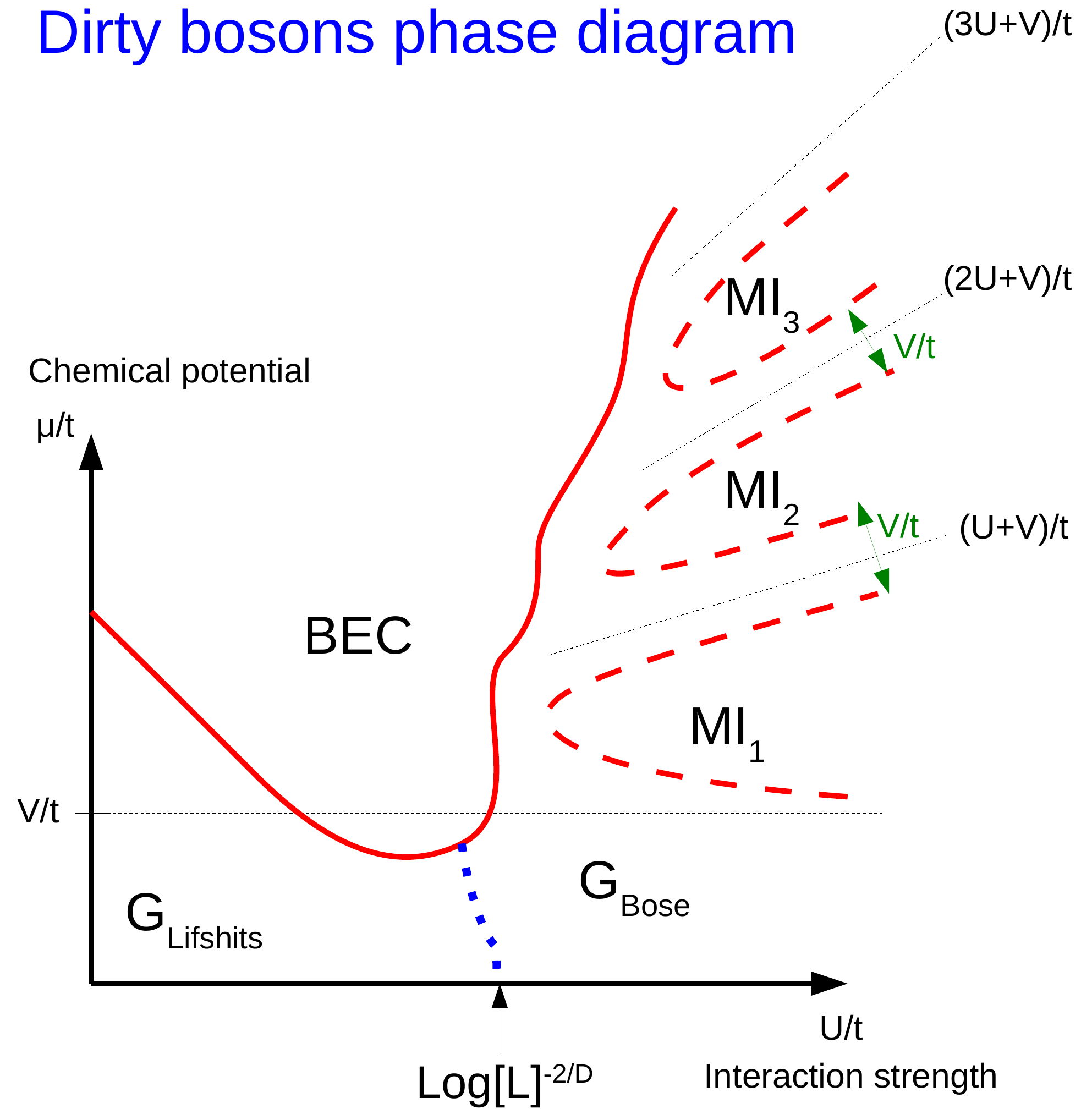}
\caption{Schematic phase diagram for dirty bosons \revision{at zero temperature}
\revision{as a function of interactions ($U$) and chemical potential ($\mu$)}
containing glassy (G), superfluid (BEC), and Mott insulating (MI) phases. The
continuous and dashed lines mark respectively the
superfluid-insulator and glass-Mott transition, while the dotted
line indicates a crossover from the Lifshits glass (heavily
fragmented) to the Bose glass (extended) region. Outside of the
BEC region, the one-body density matrix $\rho({\bf r},{\bf r'})$
decays as $|{\bf r-r'}|\to \infty$. The Mott insulator lobes
appear at sufficiently strong interactions, as the inclusion
theorem guarantees that there may be no MI for $U<V$. }
\label{fig:dirtyPhaseDiag}
\end{figure}
%%%%%%%%%%%%%%%%%%%%%%%%%%%%%%%%%%%%%%%%%%%%%%%%

In the past
decades there appeared an enormous literature on disordered systems, see,
e.g.,
\cite{Giamarchi88,Fisher89,Scalettar91,Damski03,Lugan07,Lugan07bis,Altman08,Pollet09,Sanchez-Palencia10,Gaul11,Lewenstein12},
although due to the complexity of the topic mathematically
rigorous results are still scarce.  A great deal of work has been
done on random Schr\"{o}dinger operators, which describe single
particle behaviour, see \cite{KirschSurvey} for an excellent
survey. In the recent paper \cite{Bishop11} two of us have
provided rigorous tight bounds on the ground state energy, as well
as approximations of the ground state and excited states wave-functions for the
case of random impurity (Bernoulli) disorder in a 1D lattice.
 Since in this case the shape of the low-energy states is known, there are analytical ways to estimate interaction energy and obtain rigorous results also for interacting systems.

A challenging and still open problem is the understanding of
interaction effects on disordered systems. A crucial step forward in our understanding of dirty bosonic
systems was made by Giamarchi and Schulz \cite{Giamarchi88} and
Fisher {\it et al.}~\cite{Fisher89} who conjectured that, in
presence of weak disorder, interactions give rise to an
intermediate compressible but insulating phase, the Bose glass, in
between the superfluid and the insulating strongly-interacting
phases (Tonks gas in continuous 1D systems, or Mott insulator in
lattice systems). The disputed controversy, as to whether or not
the insulating Mott phase is always surrounded by a glassy phase
was finally settled in \cite{Pollet09}, who demonstrated, using
the theorem of inclusions, that this is indeed the case for the
dirty boson system with generic bounded disorder. Experimentally,
this problem has been recently addressed in
\cite{Fallani07,Pasienski10,Gadway11}.

In the regime where disorder dominates over interactions, it was
shown instead that the ground state of the quantum fluid may be
described in terms of a qualitatively different glassy phase,
called Lifshits (or Anderson) glass \cite{Lugan07,Deissler10},
characterized by \revision{exponentially localized} and well separated "islands". The
latter may be identified with the low-lying single-particle
eigenstates residing in the regions of space where the potential
is small. As repulsive interactions are strengthened, an increasing
number of islands is populated, until their overlap becomes
sufficient to establish phase coherence and transport between
them. The gas as a whole then undergoes a phase transition towards
an extended Bose--Einstein condensate (BEC). If the gas is confined
in a lattice, for even larger interactions the repulsion between
particles becomes so strong that on-site density fluctuations
characteristic of the superfluid state become energetically
unfavourable. In this situation  the gas undergoes a further
transition out of the BEC, first to the Bose glass phase and then
into the incompressible Mott state. Between the Lifshits and the
Bose glasses there is no phase transition, as both are gapless,
insulating and compressible phases, but nonetheless the two states
are qualitatively different, in the sense that in the former the
gas is \revision{fragmented} into independent and distant low-energy islands
(the Lifshits states), while the latter tends to extend over a
large portion of the available volume. The phases discussed above
are sketched in figure \ref{fig:dirtyPhaseDiag}, and their most
important properties listed in table \ref{table:states}.
%%%%%%%%%%%%%%%%%%%%%%%%%%%%%%%%%%%%%%%%%%%%%%%%
\begin{table}[b]
\caption{Summary of most common phases for dirty bosons in a lattice, and
associated properties.} \label{table:states}
\begin{indented}
\item[]
\begin{tabular}{|c|c|c|c|c|c|}
\hline
 \multicolumn{2}{|c|}{} & Superfluid & Compressible & Gapless & \revision{Fragmented}\\ \hline\hline
 \multicolumn{2}{|c|}{BEC} & Y & Y & Y & N\\ \hline  \hline
 Glass & Lifshits & N & Y & Y & Y \\ \cline{2-6}
       & Bose & N & Y & Y & N \\ \hline \hline
 \multicolumn{2}{|c|}{Mott insulator} & N & N & N & N\\ \hline
\end{tabular}
\end{indented}
\end{table}
%%%%%%%%%%%%%%%%%%%%%%%%%%%%%%%%%%%%%%%%%%%%%%%%

In the present paper we consider disorder of Bernoulli type, i.e.,
randomly-distributed, identical and localized impurities, and we
focus on the regime of weak to intermediate interactions in
presence of such impurities. We first investigate the properties
of the ideal gas, discussing the typical size and energy of the
localized islands, and showing that the low-energy states indeed
form a so-called Lifshits tail.

We further explore interaction effects on the ground state
properties by comparing the performance of two theoretical
methods. The first one is a simple version of the multi-orbital
Hartree--Fock method (sMOHF), a variational method based on an
expansion on single-particle states. This approximation is
appropriate to describe the glassy regime where superfluidity is
suppressed by disorder~\cite{Lugan07}. More elaborate versions of
the Hartree--Fock method, that incorporate self-consistency and
allow to describe fragmentation and quantum dynamics of
interacting Bose systems can be found in \cite{Cederbaum} and
references therein. The second method is the standard
Gross--Pitaevskii (GP) approach, which \revision{generally yields
an appropriate description for every interaction strength.}

We calculate the ground state energy of the system, the associated
superfluid fraction, the fractal dimension, and introduce a new
parameter, the {\it fractional occupation}, allowing one to
discern between the Lifshits and the Bose glass. Finally, we turn
to the investigation of temperature effects in an ideal gas. Due
to the increase of kinetic energy, temperature yields effects
similar to those generated by repulsive interactions, in the sense
that the gas occupies a larger number of islands, which
increasingly overlap until the ground state covers a large amount
of the available space. Nonetheless, we will point out that there
are important differences between the two cases.

As an important result, in this paper we show that disordered
systems with Bernoulli potential allow for analytical estimates
which are very well supported by numerical simulations. These
results provided us with intuitions on how to  generalize the
rigorous results of \cite{Bishop11} to non-interacting 2D systems,
and, at least partially,  to interacting 1D and 2D systems. These
rigorous results go beyond the scope of the present article, and
will be published in a more appropriate mathematical physics
journal.

This paper is structured as follows. We introduce the disordered
potential under study and the relevant Hamiltonian in
Sec.~\ref{sec:model}. In Sec.~\ref{sec:noninteracting} we identify
the eigenstates of the ideal system, we numerically demonstrate
the Lifshits tail behaviour, and we discuss the ground state energy
scaling as a function of the size of the system. In Sec.
\ref{sec:interacting} we introduce the sMOHF and GP methods, and
in Sec.~\ref{sec:comparison} we compute a number of relevant
quantities for interacting systems. In Sec.~\ref{sec:temp} we
compare the effects of interactions and  non-zero temperature, and
we present our conclusions in Sec. \ref{sec:conclusions}.

% $$$ $$$$ $$$$ $$$$ $$$$ $$$$ $$$$ $$$
\section{The model}\label{sec:model}
We study the properties of a bosonic gas on disordered 1D and 2D
square lattices of linear dimension $L$. We consider disorder of
the so-called Bernoulli type, i.e., the potential on each site is
an independent random variable that assumes the value:
\beq V_i=\left\{
\begin{array}{ll}
  V>0 & {\rm with\,probability\,} 1-p \\
  0 & {\rm with\,probability\,} p \\
\end{array}
\right. \eeq
This kind of potential may be ideally realized by using a
two-component gas, and a species-selective lattice which deeply
traps only one of the two components
\cite{Gavish05Massignan06,Gadway11}. At sufficiently low energies,
the component which is not trapped experiences $s$-wave collisions
against localized scatterers. Alternatively, a Bernoulli disorder
may be imprinted directly on the gas exploiting single-site addressable
optical lattices \cite{SingleSite}.

The use of Bernoulli disorder is particularly appealing
because its asymptotic properties, as discussed in the next sections, become
apparent even for small lattices (e.g., with $50^D$ sites). As we show in
 \ref{app:convergence}, the Bernoulli potential has actually optimal
properties of convergence. Therefore, the Bernoulli disorder is
suitable for numerical simulations and due to its simple form, it
allows for various analytical estimates.

The Hamiltonian of the interacting system we consider is then
\beq \hat{H}=-t\sum_{\langle i,j \rangle}\hat{c}_i^\dagger\hat{c}_j+\sum_i
V_i\hat{c}_i^\dagger\hat{c}_i+\hat{U}_{\rm int}, \eeq
where $t$ denotes the tunnelling energy, $\hat{c}_i^\dagger$ is the
creation operator of bosons on site $i$, and $\langle i,j \rangle$
denotes nearest neighbour pairs of sites. The interaction term
$\hat{U}_{\rm int}$ will be discussed later in Section
\ref{sec:interacting}.

% $$$ $$$$ $$$$ $$$$ $$$$ $$$$ $$$$ $$$
\section{Properties of a non-interacting gas}\label{sec:noninteracting}

Let us start by discussing the properties of a non-interacting
system, and in particular look at its ground state energy, and the
density of low energy excited states.

\subsection{Ground state energy}
In a large 1D system with Bernoulli disorder, the linear size
$L_{\rm max}^{\rm (1D)}$ of the largest island of contiguous
zero-potential sites scales as $L_{\rm max}^{\rm (1D)}\propto \log
L$ \cite{Bishop11}. The proof of this fact is based on the
following simple observation. Let $l_0$ be the characteristic size
of the largest island, then $p^{l_0}$ estimates its occurrence.
The number of islands of similar size is expected to be of order of
$L$ (strictly speaking $L/\log{L}$ as we shall discuss below),
so that the probability of occurring of any one of them is
$Lp^{l_0} \to {\rm const}$ as $L\to\infty$, which indeed implies
$l_0 \simeq \log{L}/|\log(p)|$. It is then clear that the ground
state energy of the non-interacting bosonic gas in such a potential
will be bounded above by the energy of the first harmonic, i.e., a
half-sine function, of the largest island. A lower bound of the
same order was proven in \cite{Bishop11}, showing that the ground
state energy behaves asymptotically as $\pi^2/(\log L)^2$ when $L
\to \infty$.

Similarly, one can argue that in a large D-dimensional system the
ground state will be occupying the largest spherical island of
\revision{zero-potential} sites since this shape minimizes the
kinetic energy \cite{Antal95}. Its diameter can be shown to grow
as $L_{\rm max}^{\rm (D)}\propto (\log L)^{1/D}$, and therefore we
expect the energy of the ground state to scale as $E_0\propto \log
L^{-2/D}$. Again, the proof is based on the similar argument as in
the 1D case, except that one has to replace the characteristic
length $l_0$ by the volume $l_0^D$ in arbitrary dimension. The
numerical confirmation of the scaling for the 2D case is shown in
the inset of figure \ref{fig:LifshitsTailAndGroundStateScaling}.

 For comparison, we plot in the same inset the ground state energy obtained from a
random-amplitude disorder (i.e., one where to each site
corresponds a random potential $V_i$, uniformly distributed in the
interval $[0,V^{'}]$). We see that the rate of convergence of the
energy for a Bernoulli distribution is quicker than for a uniform
distribution. Because of this, the potential with a Bernoulli
distribution is ideal for ground state energy convergence as well
as localization of low energy states. This may be quickly seen as
follows, while further details are given in \ref{app:convergence}.
If the potential takes value zero with a positive probability $p$,
one can define islands of zero potential as the natural locus of
the ground eigenstate.   In cases where the values of the
potential are positive, even when they are arbitrarily close to
zero, the low potential islands can still be defined, using a
(volume-dependent) energy cut-off, but the contribution to energy
from such islands will in this case be larger.  Accordingly, the
convergence of the ground state energy to zero will be faster in
the former case.  Among the distributions which assign a fixed
probability $p$ to the zero value of the potential, the optimal
ones to work with are the Bernoulli distributions, in which some
positive value $V$ is assumed with probability $1-p$.  While the
rate of convergence of the ground state energy to zero is
comparable for all such distributions, the advantage of the
Bernoulli distribution is that it localizes the low energy states
more clearly.  The analysis of other potential distributions is
more complicated, because it necessitates introducing an energy
cut-off to define low potential islands.  The Bernoulli potential
is also easier to work with numerically, since it requires
sampling fewer potential realizations.

Although we do not expect any of the properties discussed in the
following to depend on the specific type of bounded disorder, we
clearly see that the Bernoulli choice yields a much faster
convergence of the ground state energy to the desired asymptotic
behaviour.

\subsection{Lifshits tail}

%%%%%%%%%%%%%%%%%%%%%%%%%%%%%%%%%%%%%%%%%%%%%%%%
\begin{figure}
\centering
\includegraphics[width=0.5\textwidth]{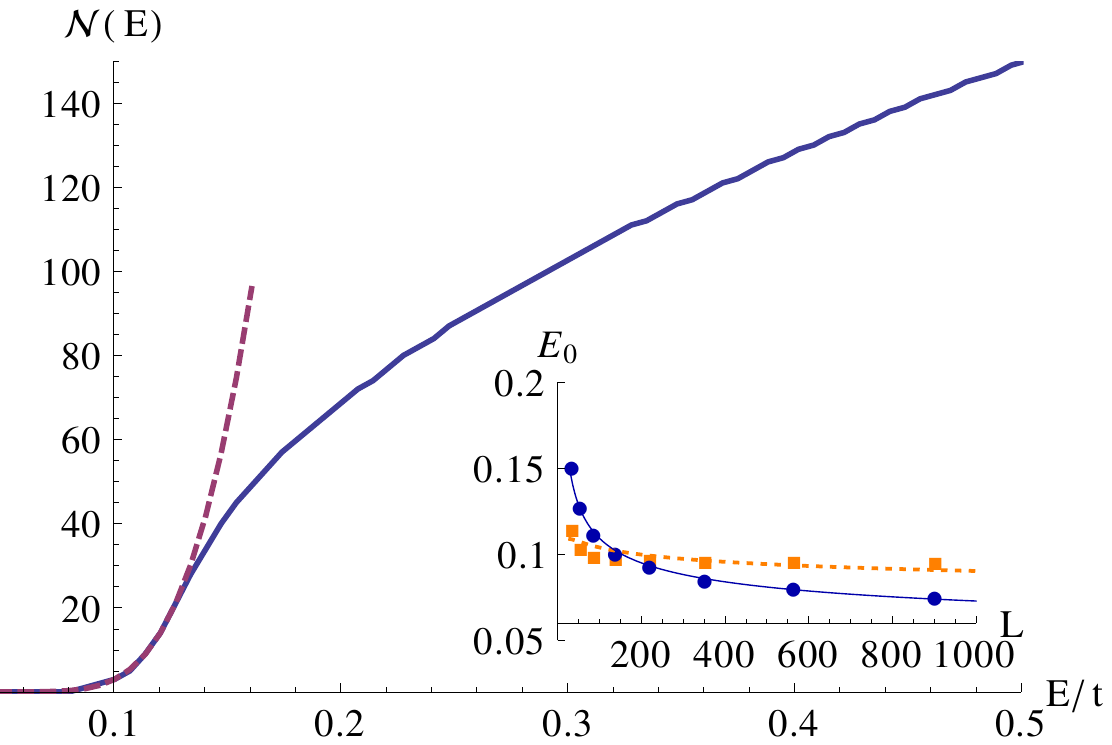}
\caption{Properties of non-interacting dirty bosons in 2D. Main
figure: the cumulative density of states $\mathcal N(\epsilon)$
shows a Lifshits tail (purple dashed line) given
by\ (\ref{eq:LifshitsTail}). Here we have taken a
200$\times$200 lattice, and averaged over 10 realizations. Inset:
ground state energy $E_0$ in 2D plotted versus linear size of the system
$L$; results for Bernoulli disorder are plotted in blue and scale as $\sim1/\log
L$, while results for random-amplitude disorder are plotted in orange and scale
as $\sim \log \log L/\log L$.} \label{fig:LifshitsTailAndGroundStateScaling}
\end{figure}
%%%%%%%%%%%%%%%%%%%%%%%%%%%%%%%%%%%%%%%%%%%%%%%%
Since the disordered potential is chosen to be finite, the spectrum of the system is bounded from below. As a
consequence, the spectrum is expected to exhibit a so-called
Lifshits tail \cite{Lugan07,Simon85},  in the sense that
the cumulative density of states (cDOS) $\mathcal
N(\epsilon)=\int^\epsilon\mathrm{d}\epsilon'\
\mathcal{D}(\epsilon')$ behaves at low energies as
\beq \mathcal{N}(\epsilon) \sim \exp
\left[-c(\epsilon-V_{\rm min})^{-D/2}\right].
\label{eq:LifshitsTail} \eeq
The cDOS and the associated Lifshits tail for the considered  2D
system were obtained numerically and are shown in figure
\ref{fig:LifshitsTailAndGroundStateScaling}. The Lifshits states
lie on well-separated islands, and have almost degenerate energies
since the islands' diameters are approximately the same.

% $$$ $$$$ $$$$ $$$$ $$$$ $$$$ $$$$ $$$
\section{Interacting case}\label{sec:interacting}
In this Section we will present two common theoretical approaches
used to describe a Bose gas with short-range interaction
potential. The first one corresponds to a simplified version of
Multi-Orbital Hartree--Fock (sMOHF) method, based on an expansion
into single-particle eigenstates. As we will see, sMOHF is
suitable to describe a weakly-interacting system, whose ground
state occupies few disconnected large islands of zero potential.
For more sophisticated version of the Hartree--Fock approach
applied to bosons, see \cite{Cederbaum}. The second method is the
standard Gross--Pitaevskii (GP) equation, which describes the
dynamics of a fully-coherent matter wave, and \revision{correctly
describes also the regime of strong interactions, where large
overlaps between the islands and self-interaction on
each island play an important role.} overlap between the islands.

Before going into details, let us estimate when the sMOHF
description based on non-interacting single particle states should
be valid. As we have discussed, this regime corresponds to the
Lifshits glass region. Let us assume that we have $\tilde r$ {\it
filled} islands of the characteristic size $l_0$, and "volume"
$l_0^D$, so that
$$\tilde rl_0^D=CL^D,$$
where  the proportionality constant $C$ depends,
in general, weakly on $p$ and the density $\rho$,  since it
results from the complex interplay between the kinetic and
interaction energy. In the following we will ignore this
dependence.

The Lifshits glass regime occurs then for an inter-particle
interaction coupling $g$ smaller than the characteristic value
$g_{\rm ch}$, at which the kinetic energy is comparable to the
interaction energy. An estimate of the characteristic coupling is
easily obtained by equating the kinetic energy per particle
$\simeq 1/l_0^2$, with the interaction energy $\simeq gN/\tilde
rl_0^D$. Using the above expression for $\tilde{r}$, we obtain
\beq g_{\rm ch}\simeq C/\rho l_0^2 \simeq
C(|\log{p}|)^{2/D}/\rho(|\log{L}|)^{2/D}. \label{eq:g_ch}\eeq
\revision{This scaling provides a good estimate of the coupling at
which the two energies depicted in figure
\ref{fig:MOHFandGPenergies} start to diverge.}

\subsection{Multi-orbital Hartree--Fock approach}
In the sMOHF treatment, the wave function is expressed in the
single particle eigenstates, and is taken to be in the product
form
\beq |\Psi\rangle=\prod_k \frac{(\hat{a}^\dagger_k)^{n_k}}{\sqrt{n_k!}
}|0\rangle, \eeq
where $|0\rangle$ is the vacuum of the system and $a^\dagger_k$
creates a particle in the $k$th non-interacting eigenstate
(i.e., orbital). We consider repulsive interactions of
strength $g>0$, which yield an interaction energy given by
\begin{eqnarray}
\langle \hat{U}_{\rm int}\rangle=\frac g 2 \int \rm{d}\bx \
\hat\psi^\dagger(\bx)\hat\psi^\dagger(\bx)\hat\psi(\bx)\hat\psi(\bx)=\nonumber\\
 \frac g 2 \int \rm{d}\bx \sum\psi^*_{k_1}(\bx)\psi^*_{k_2}(\bx)\psi_{k_3}(\bx)\psi_{k_4}(\bx)\hat{a}^\dagger_{k_1}\hat{a}^\dagger_{k_2}\hat{a}_{k_3}\hat{a}_{k_4},
\end{eqnarray}
where the sum runs only over terms conserving number of particles,
i.e., such that $k_1+k_2=k_3+k_4$. The
diagonal terms ($\hat{a}^\dagger_{k}\hat{a}^\dagger_{k}\hat{a}_{k}\hat{a}_{k}$) contribute
a factor $\frac g 2 \sum_{k} {n}_k({n}_k-1)O_{k,k}$, where we have
defined the overlaps \mbox{$O_{k,j}=\int {\rm d}{\bf
x}|\psi_k({\bf x})|^2|\psi_j({\bf x})|^2$}.

There are two other possibilities which conserve the number of
particles: ($k_4=k_1$ and $k_3=k_2$), or ($k_4=k_2$ and
$k_3=k_1$). These two terms, called direct and exchange terms, are
equal for bosons, and their sum contributes a factor $\frac g 2
\sum_{j\neq k}2n_k n_j O_{k,j}$.

The average interaction energy reads then:
\beq
\langle \hat{U}_{\rm int}\rangle =\frac g 2 \sum_{k}\left[n_k(n_k-1)O_{k,k}+\sum_{j\neq k} 2n_kn_jO_{k,j}\right].
\eeq

 The occupation
probabilities $n_k$ yielding the ground state may now be found by
minimizing the total energy
\beq \label{eq:total_energy} E=\sum_k n_k E_k+\langle \hat{U}_{\rm
int}\rangle, \eeq
subject to the constraints of normalization and positivity of all $n_k$,
\beq \label{nc} \sum_k n_k=N,\quad \forall k,\, n_k\geq 0.\eeq
We present the analytic solution of this problem in
\ref{app:completeSol}.

\subsection{Gross--Pitaevskii approach}
The properties of the system may be analysed also in the framework
of the usual GP equation,
\beq \label{eq:GP} \left(-\frac{\hbar^2}{2m} \triangle+V+g N
|\psi(\bx)|^2\right)\psi(\bx)=\mu\psi(\bx) \eeq
which describes the behaviour of an interacting Bose gas in terms
of a single-particle coherent wave-function. We find the ground
state and the chemical potential $\mu$ by imaginary time
evolution, applying an operator split method. The ground state
energy $E$ is related to $\mu$ by the formula
\beq E=\mu - \frac{gN}{2} \int \rm{d}\bx\, |\psi(\bx)|^4.\eeq
\section{Comparison of $\mathbf{sMOHF}$ and GP approaches}\label{sec:comparison}

%%%%%%%%%%%%%%%%%%%%%%%%%%%%%%%%%%%%%%%%%%%%%%%%
\begin{figure}[t]
\centering
\includegraphics[width=0.48\textwidth]{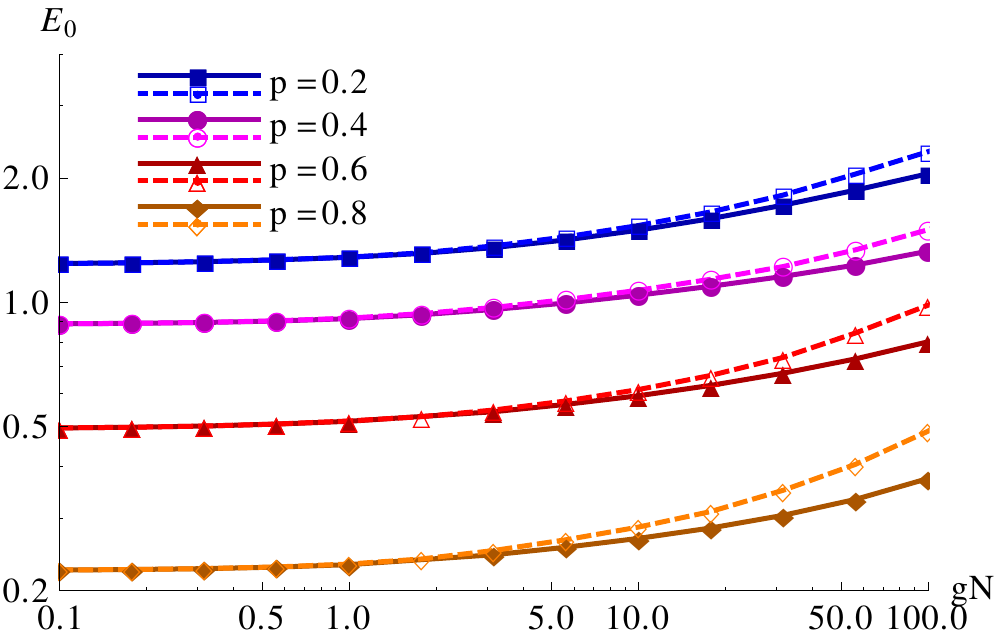}
\caption{\revision{Ground state energies obtained by sMOHF
(dashed) and GP (continuous) for various disorder densities $p$.
The value of the coupling constant at which the energies obtained
by the two methods start to diverge is estimated by equation
(\ref{eq:g_ch}).
Results obtained by averaging over 5 simulations and $V=5t$.}}
\label{fig:MOHFandGPenergies}
\end{figure}
%%%%%%%%%%%%%%%%%%%%%%%%%%%%%%%%%%%%%%%%%%%%%%%%
In this Section, we present the numerical solutions of the sMOHF
and GP equations, comparing the performance of the two methods. In
particular, we discuss and compare the ground state energy, the
superfluid fraction, and the fractal dimension. \revision{As we
will see, in the regime of strong disorder (Lifshits regime) sMOHF
provides ground state energies in accord with GP. Nonetheless,
sMOHF has the advantage of being insensitive to convergence
issues, and its \revision{solution} is generally much faster than
GP. \revision{The fact that in this regime both methods agree
confirms the intuition that the particles populate low-lying
eigenstates, the Lifshits islands}. In the regime where
interactions dominate over the disorder, we will show instead that
GP provides energies which are perceptibly lower than sMOHF.}
Unless otherwise noted, the numerics presented in this section
have been obtained \revision{for a system on a 2D lattice with
$32\times32$ sites,  $N_{\mathrm{part}}=10^4$ particles,
a Bernoulli potential with $V=5t$ or $V=50t$}, and we have set
$t=\hbar^2/2m=k_B=1$. Where needed, we have performed appropriate
averages to obtain results which are independent of the particular
disorder configuration.

\subsection{Energy}
In figure \ref{fig:MOHFandGPenergies} we compare the energies
obtained through the sMOHF approach with those obtained from the
GP equation as a function of the interaction strength $g$. For a
coupling constant $g$ smaller than a characteristic value $g_{\rm
ch}$, the \revision{minimization of expression
(\ref{eq:total_energy}) and the GP equation yield the same values
of energy proving that sMOHF method correctly describes the system
in this limit.}
The characteristic value $g_{\rm ch}$ is given by the interaction strength at
which the energy obtained from the sMOHF \revision{starts to differs from the
one obtained by the GP equation.} The
range of $g$ for which sMOHF approach is \revision{an equivalent
description} shrinks with increasing $p$,
as the scatterers become increasingly sparser, leaving large
regions of zero potential. The range of applicability of sMOHF
also shrinks with decreasing $V$. For values of the interaction
$g\gg g_{\rm ch}$, the disorder plays a negligible role and the
energy per particle saturates to the analytic value
$E=\overline{E}+g \rho/2$, where $\overline{E}$ is a constant that
depends on $V$ and $p$. This analytic result is recovered by the
numerical solution of the GP equation, but not by the sMOHF
approach.

%%%%%%%%%%%%%%%%%%%%%%%%%%%%%%%%%%%%%%%%%%%%%%%%
\begin{figure*}
a)\includegraphics[width=0.48\textwidth]{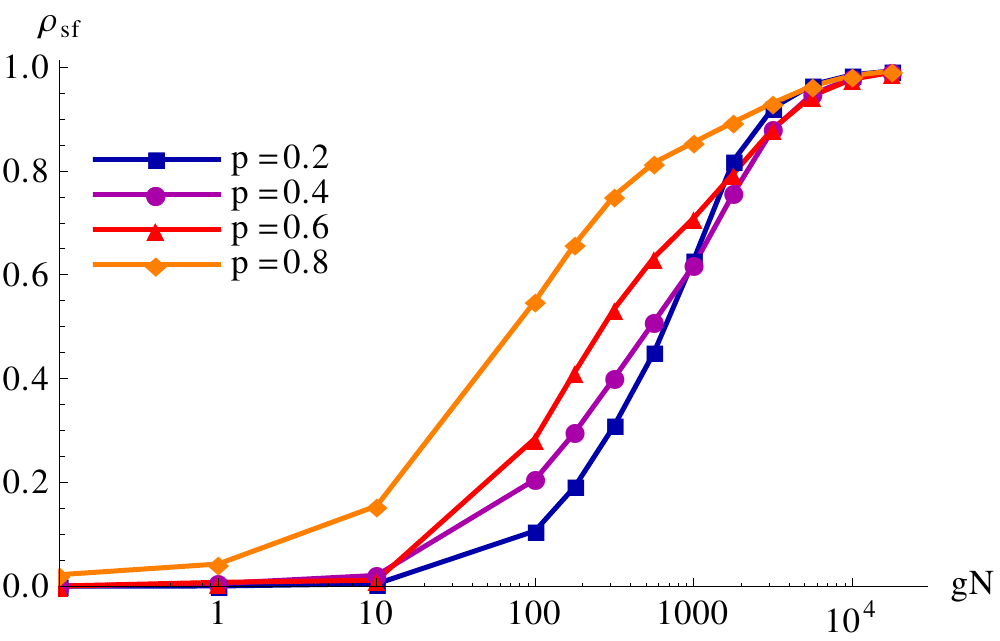}
b)\includegraphics[width=0.48\textwidth]{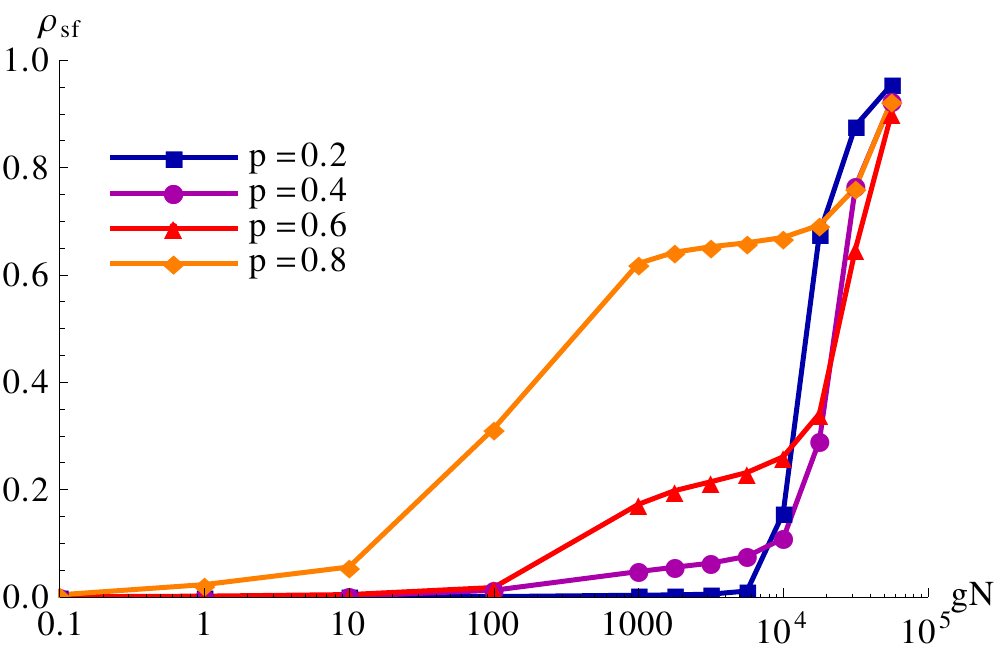}
\caption{Superfluid fraction $\rho_{\rm sf}$ as a function of
potential density $p$ and interaction strength $g$ \revision{for
a) $V=5t$ b) $V=50t$.
}
} \label{fig:SFfraction}
\end{figure*}
%%%%%%%%%%%%%%%%%%%%%%%%%%%%%%%%%%%%%%%%%%%%%%%%
\subsection{Superfluid fraction}
The superfluid fraction $\rho_{\rm sf}$ may be defined in terms of
the energy change of a periodic system in response to twisted
boundary conditions along one direction. To calculate this
quantity we solve the GP equation (\ref{eq:GP}) with boundary
conditions
 $\psi_{i+L,j}=\ee^{\ii \Phi} \psi_{i,j}$, and we obtain
\beq \rho_{\rm sf}=\frac{2mL^2}{\hbar^2} \frac{E(\Phi)-E(0)}{\Phi^2}, \eeq
where $E(\Phi)$ is the energy per particle of a system with the
total phase shift $\Phi$. \revision{Our results are depicted in figure \ref{fig:SFfraction}.
In the regime where sMOHF provides a good description, the SF fraction
is negligible because the localized single-particle eigenstates experience an exponentially small
energy change due to the imposed phase shift.
 The GP equation instead yields a perceptibly lower energy than sMOHF when the
 superfluid fraction $\rho_{\rm sf}$ becomes sufficiently large ($ \rho_{\rm sf}\gtrsim 0.1$).}

%%%%%%%%%%%%%%%%%%%%%%%%%%%%%%%%%%%%%%%%%%%%%%%%
\begin{figure}
\centering
\includegraphics[width=0.5\textwidth]{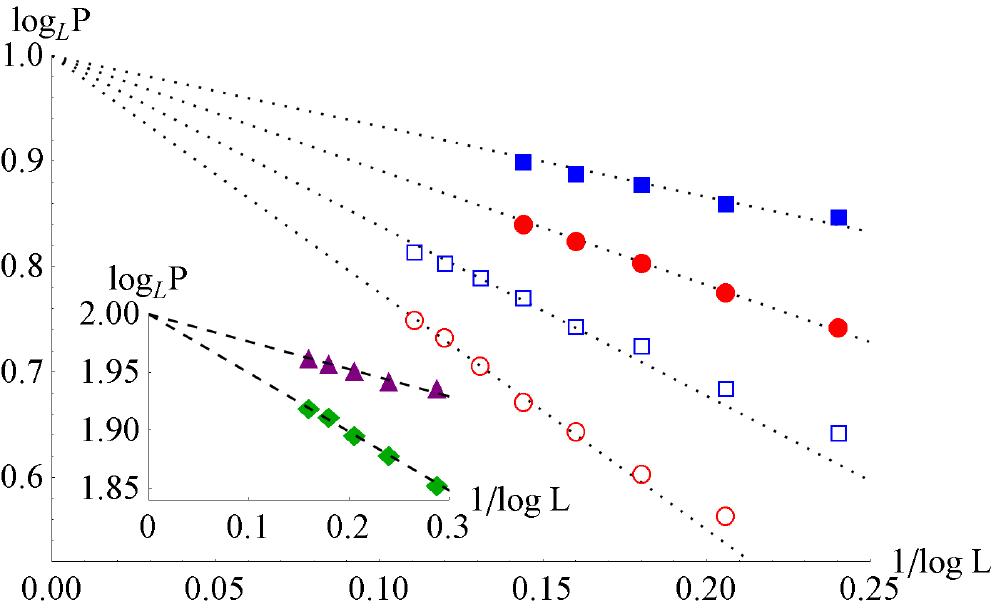}
\caption{Scaling of the logarithm of the participation number
$\log_L P$ versus the inverse of the logarithm of the lattice
size. The slope of the lines, being the logarithm of $c$, allows
for determining the fraction of space occupied by the state. Main
figure: 1D lattice of length $L$, and interaction strengths
$g=0.1$(empty) and $g=10$(filled). The potential densities are
respectively $p=0.4$(red circles) and $0.7$(blue squares). Inset:
2D lattice of linear size $L$, interaction strength $g=10$,
potential parameter $p=0.6$(green diamonds), $0.8$ (purple
triangles).}\label{fig:fracDimvsL}
\end{figure}
%%%%%%%%%%%%%%%%%%%%%%%%%%%%%%%%%%%%%%%%%%%%%%%%

\subsection{Fractal dimension}
As we have discussed in the previous sections, the ground state of
a weakly-interacting fluid is localized on one or a few Lifshits
islands. When interactions start to play an important role, the
extension of the ground state increases generally until the gas
occupies all available space. In order to provide a quantitative
measure of the extension of the ground state, we calculate its
fractal dimension $d^*$ \cite{Kramer93}, and introduce here the
concept of \emph{fractional occupation} $c$.

 The fractal dimension is defined as a minimum $d^{*}$ such that
\beq \lim_{L\rightarrow\infty} \frac{P}{L^{d^*}}=c,\, c>0.
\label{eq:fracdim_def}\eeq

Here $P=1/\int \rm{d}\bx |\psi_0(\bx)|^4=1/O_{0,0}$ is the
so-called ``participation number'' of the ground state
\cite{Kramer93}. For an extended state such as a plane wave one
finds that $P$ is proportional to the volume of the system, and
therefore the fractal dimension equals the Euclidean
dimension, $d^*= D$. For a state which is instead localized in a
volume $Q$, the participation number behaves as $Q$, and the
corresponding fractal dimension vanishes if $Q$ grows slower than
any power of $L$. For instance, the ground state of the
non-interacting system is localized on an island of volume $Q\sim
\log L$ which for all $\alpha$ grows slower then $L^{\alpha}$. So,
in general, one has $0\leq d^*\leq D$.

Analysing the results obtained from both sMOHF and GP approaches, we find
that $d^*=D$ for any $g>0$. This may be seen as follows.  Assuming
that $d^*$ is bounded strictly below $D$, we see that the
interaction energy $(gN/2)\int \rm{d}\bx\, |\psi_0(\bx)|^4$ will
diverge. Physically, there is not enough space for the particles
to keep the interaction energy bounded unless they spread out
through a non-zero proportion of the whole space. If $\log_{L} P =
D - f(L)$, and $\lim_{L\to\infty} L^D/L^{d^*} = c$, then it must
be that $L^{f(L)} \to c$, so $f(L) \approx \log(c)/\log(L)$.
Numerical results confirm this behaviour in both 1D and 2D.
To show this, in figure \ref{fig:fracDimvsL} we plot $\log_L P$ as
a function of $1/\log L$ for a 1D system and a 2D system (inset).

%%%%%%%%%%%%%%%%%%%%%%%%%%%%%%%%%%%%%%%%%%%%%%%%
\begin{figure*}
a)\includegraphics[width=0.48\textwidth]{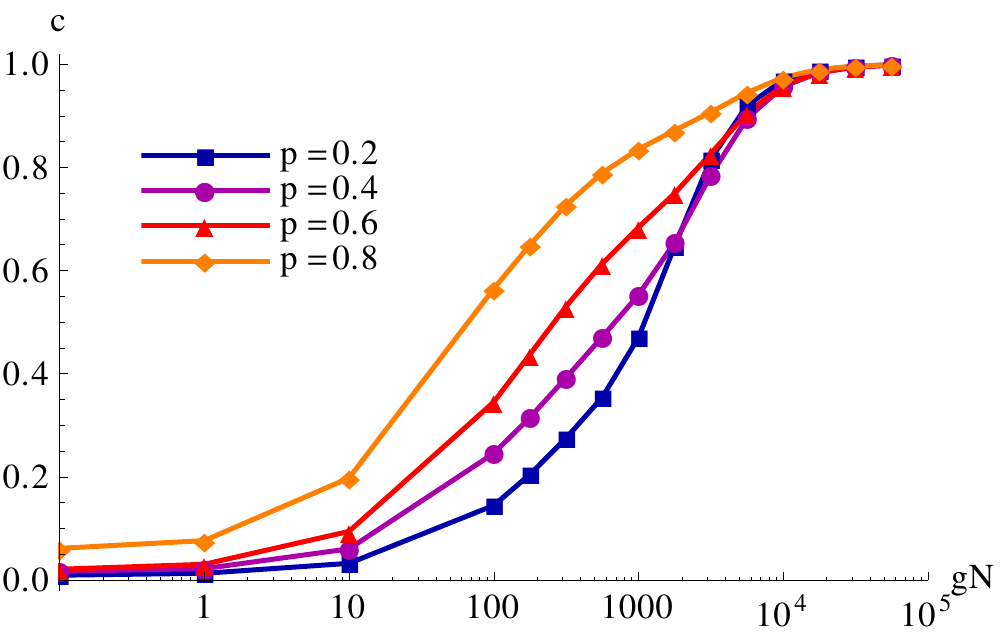}
b)\includegraphics[width=0.48\textwidth]{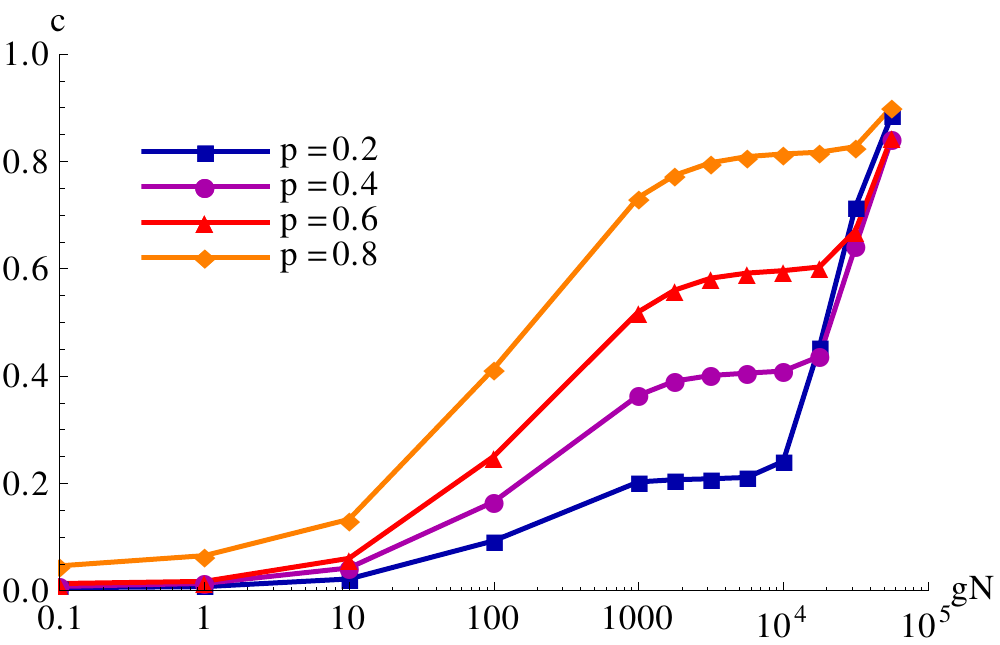}
\caption{Fractional occupation $c$ of the ground state as a
function of potential density $p$ and interaction strength $g$ for
\revision{a) $V=5t$ and b) $V=50t$}. Here we have considered a 2D
lattice and $c$ was obtained by averaging over 5 potential
realizations.} \label{fig:c_vs_g_vs_p}
\end{figure*}
%%%%%%%%%%%%%%%%%%%%%%%%%%%%%%%%%%%%%%%%%%%%%%%%

The quantity $c$, which takes values between 0 and 1, can be
interpreted as the fraction of space occupied by the ground state,
and we will therefore refer to it as the \emph{fractional
occupation} of the ground state. We show our results for the
fractional occupation in figure \ref{fig:c_vs_g_vs_p}.

\revision{For strong disorder ($V\gg t$)} the fractional
occupation $c$ displays at intermediate interactions a plateau at
$\sim p$. \revision{This coincides with a similar plateau in the
plot of the superfluid fraction (cf.\ figures \ref{fig:SFfraction}b and \ref{fig:c_vs_g_vs_p}b).} In
1D the explanation for such behaviour of $c$ \revision{and
$\rho_{\mathrm{sf}}$} is the following. For small $g$ the wave
function fills only a part of zero-potential space. Indeed, for
$g=0$, we have a linear operator and therefore the ground state is
approximately the wave function on the longest island of zero
potential, yielding a nearly zero $c$. For $g$ small such that
$gN/2$ is approximately $1/(\log L)^{2}$, the ground state spreads
to many of the long islands: the kinetic energy increases a little
(on the order $1/(\log L)^3$) while interaction energy decreases
by a factor $1/(\#$ of islands). As $g$ grows to be on the order
of $1$, it supports itself on all sites of zero potential save for
the shortest islands, so $c \approx p$. This is where the graphs
level off in (Fig \ref{fig:c_vs_g_vs_p}), where the ground state
does not have large norm on short islands of zero potential and
sites of $V$ potential.

For large interactions one may use the Thomas-Fermi Approximation,
cf. \cite{Lewenstein12}. Suppose we choose an ansatz such that the
wave function equals $\frac{m}{\sqrt{pL}}$ on sites of zero
potential and $\frac{\sqrt{1-m^2}}{\sqrt{qL}}$, then energy
optimizing $m$ would be $m^2 = p + \frac{pqV}{g\rho}$ and $ 1 -
m^2 = q - \frac{pqV}{g\rho}$. If $g \rho \leq p V$, then the above
ansatz is invalid, and the ground state stays exclusively on sites
of zero potential. If we compare the energy of a wave function
equal to $\frac{1}{\sqrt{pL}}$ on sites of zero potential and zero
elsewhere to a wave function that is equal to $\frac{1}{\sqrt{L}}$
everywhere, and ignore the kinetic energy terms (which we can
control by another parameter), we see that the first wave function
provides a better variational ansatz  if $\frac{g\rho}{2} \leq
pV$. This means that for $g\rho \lesssim 2pV$, we are in an area
where $c \approx p$, that is, the kinetic energy and interaction
energy are balancing, but any significant support of the ground
state on the sites occupied by scatterers is too costly.
\revision{Since the wave function does not spread significantly in
this regime, also the superfluid fraction remains constant.}

\section{Comparison of temperature and interaction effects}\label{sec:temp}

In this section we turn to the analysis of the effects of
non-zero temperature in a non-interacting system. We will see
that those remind very much effects of interactions at zero
temperature, and yet are quite different.

\subsection{Single-particle density distribution}
For free boson systems, the effects of temperature may be studied
by expanding arbitrary states in the basis of
non-interacting eigenstates, in analogy with what we did to treat
the interacting case. In this approximation, the average density
of the gas reads
\beq \rho({\bf x})=L^{-D}\sum n_k |\psi_k({\bf x})|^2. \eeq
An approximation of a typical wave-function is then given by
\beq \psi({\bf x})=\sum \sqrt{w_k}{\rm e}^{i\phi_k}\psi_k({\bf x}), \eeq
where ${\rm e}^{i\phi_k}$ are random phases, and $w_k$ is a Gaussian random
variable with distribution $\mathcal{P}(w_k)=n_k^{-1}{\rm e}^{-w_k/n_k}$, such
that $\langle w_k\rangle_{\mathcal{P}}=n_k$.

The occupation factors depend on the statistics of the particles.
Identical bosons follow the Bose--Einstein distribution \beq
n_k=\frac{1}{{\rm e}^{(E_k-\mu)/T}-1}. \eeq

Distinguishable (or classical) particles would instead populate the
non-interacting eigenstates following a Boltzmann distribution, $n_k= {\rm
e}^{-(E_k-\mu)/T}$.

At zero temperature, only the lowest energy state will have a
non-zero population, i.e., $n_k=N\delta_{0,k}$. Non-zero
temperatures will modify the occupation probabilities,
redistributing population over the higher energy states, but we
expect that for $T\ll t$ only the lowest energy states (the
Lifshits states) will be populated.

\revision{We note here that in an interacting 2D Bose gas of finite extension one
generally expects a crossover between two qualitatively different
regimes. At very low temperatures ($T<T_C$) phase correlations in the gas decay algebraically with distance. For
temperatures larger than the critical value $T_C$ instead, phase
correlations decay exponentially with distance. This switch in the decay of correlations can be traced back to the dissociation of bound vortex-antivortex pairs at $T>T_C$. The crossover
becomes a real phase transition in an infinite system, and the
underlying mechanism goes under the name of Berezinskii-Kosterlitz-Thouless
(BKT) transition \cite{BKT}.

Our approach to finite temperatures is rather crude, and is not
well suited to describe BKT physics, which requires considering
both temperature and interactions at the same time. However, for
the range of parameters considered in this manuscript, our model
provides a qualitative, although simplified, explanation of the
complementary roles payed by temperature and interactions.
Moreover, a similar scenario to the one considered here is present
in three dimensions, where a true BEC exists at low temperatures.}

 %%%%%%%%%%%%%%%%%%%%%%%%%%%%%%%%%%%%%%%%%%%%%%%%
\begin{figure*}
\centering
a)\includegraphics[width=\textwidth]{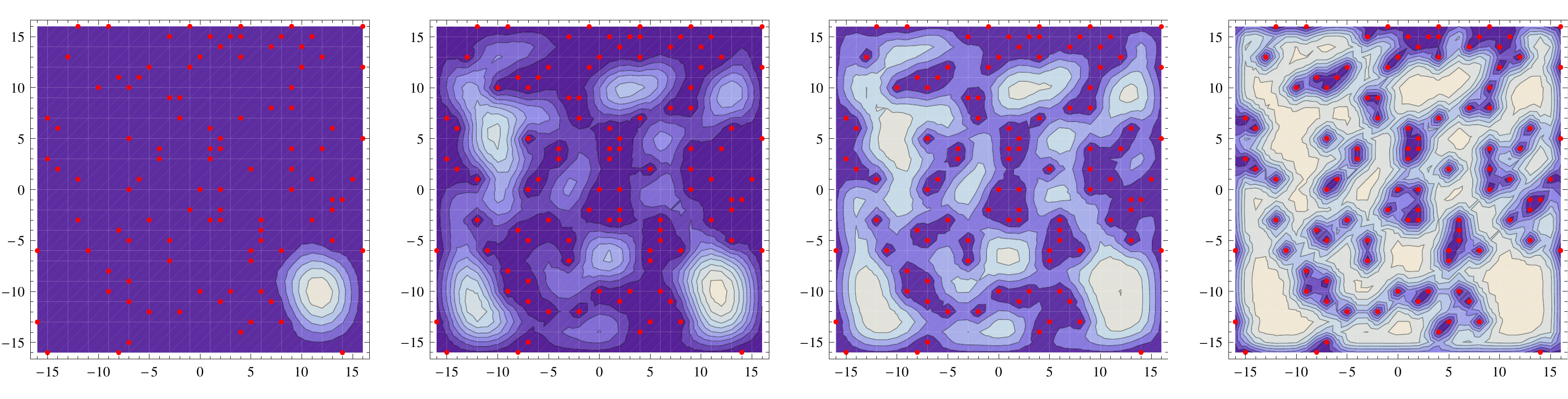}
b)\includegraphics[width=\textwidth]{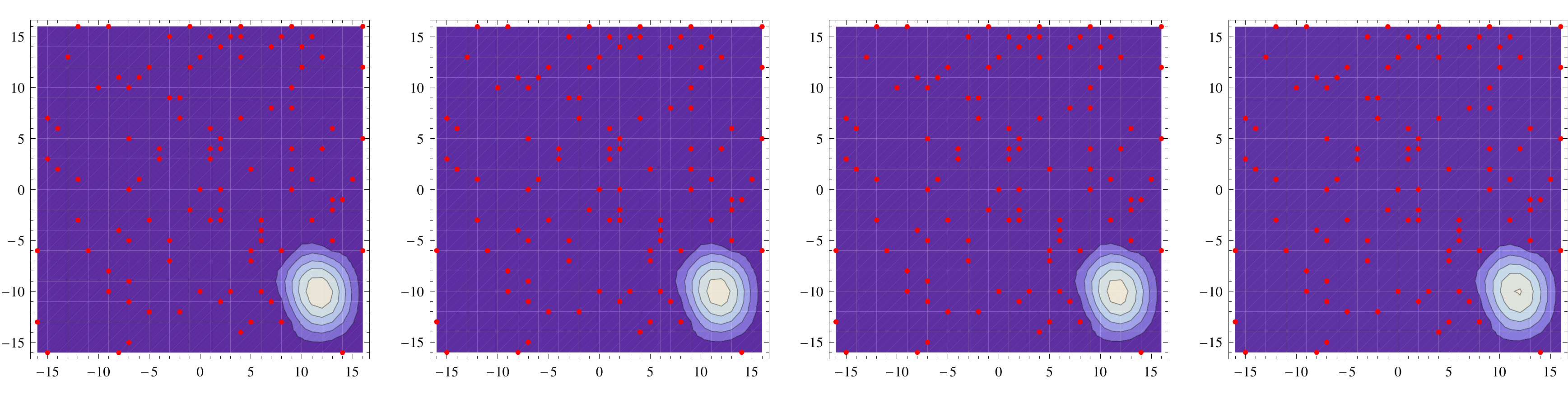}
c)\includegraphics[width=\textwidth]{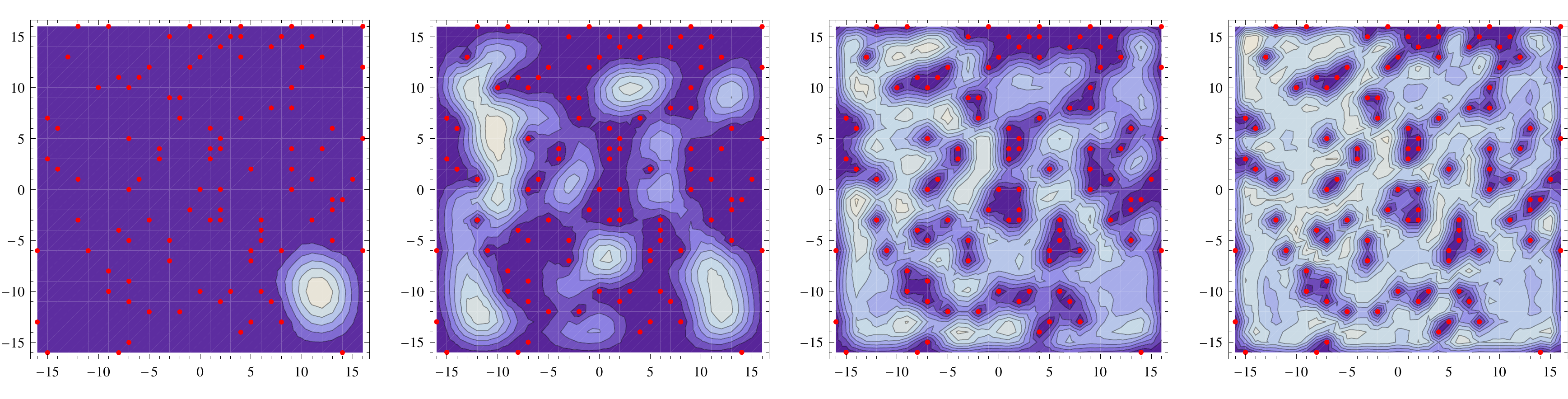} \caption{Comparison
of temperature and interaction effects. First two rows: thermally
averaged single particle density with respectively Boltzmann (top)
and Bose (centre) statistics; from left to right, $T/t=$0, 0.1,
0.2, 0.5. \newline Bottom row: single particle density of an
interacting gas; from left to right, $g\rho/t=0,\ 0.1,\ 2,\
10.$\newline The red dots indicate the sites occupied by the
disordered scatterers. The average density is
$\rho=0.9$particles/site.}
\label{fig:temperatureAndInteractionEffects}
\end{figure*}
%%%%%%%%%%%%%%%%%%%%%%%%%%%%%%%%%%%%%%%%%%%%%%%%
%%%%%%%%%%%%%%%%%%%%%%%%%%%%%%%%%%%%%%%%%%%%%%%%
\begin{figure*}
\centering
\includegraphics[width=\textwidth]{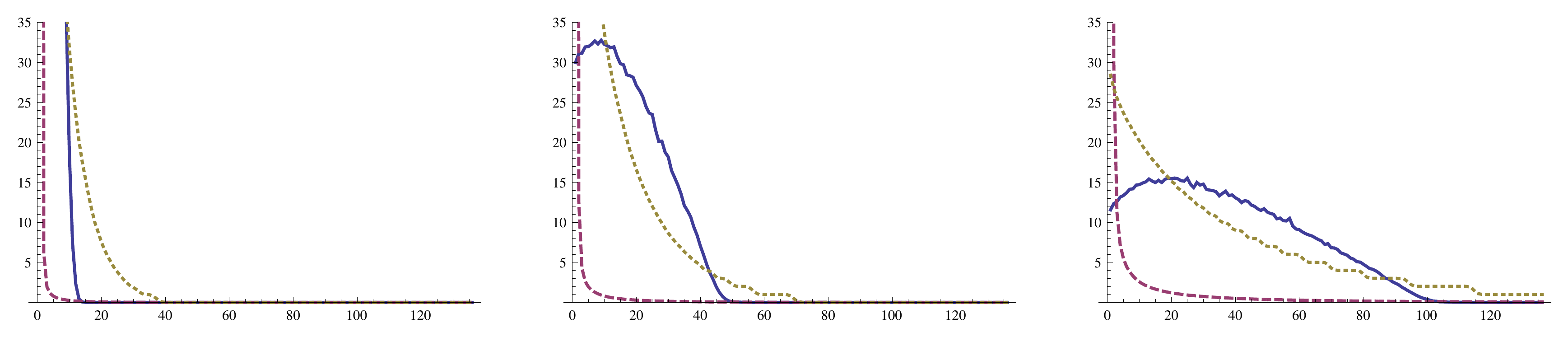}
\caption{Occupations of the single particle states for the
simulations shown in columns 2,3, and 4 of
figure \ref{fig:temperatureAndInteractionEffects}. The lines are:
Boltzmann (yellow dotted), Bose (red dashed), and sMOHF (blue
continuous). Results averaged over 200 simulations.
\revision{x-axis: index labeling the single-particle state; y-axis: occupation.}}
\label{fig:occupations}
\end{figure*}
%%%%%%%%%%%%%%%%%%%%%%%%%%%%%%%%%%%%%%%%%%%%%%%%

% $$$ $$$$ $$$$ $$$$ $$$$ $$$$ $$$$ $$$
\subsection{Non-zero temperature}
As temperature is raised, the cloud gets to occupy more and more
islands; when $T\lesssim V$ it will populate all unoccupied sites,
and for $T\gg V$ it will occupy even sites inhabited by
scatterers.

The averaged densities in presence of positive non-zero
temperature are shown in
figure \ref{fig:temperatureAndInteractionEffects}a) and b). While a
classical cloud quickly spreads over various low energy states, a
bosonic gas at low temperatures tends to condense in a single or
few Lifshits states. Interaction effects in a bosonic cloud at
$T=0$, instead, yield at first sight a state which very much
resembles the Boltzmann case. Even for moderate repulsive
interactions, the cloud spreads over various Lifshits states.
Effects of interactions at $T=0$ are shown in
figure \ref{fig:temperatureAndInteractionEffects}c.

The occupation probabilities $n_k$ for the three cases analysed
above (Boltzmann, Bose, and interacting) are shown in figure
\ref{fig:occupations}. Since the lowest-energy single-particle
states correspond to the localized and well separated Lifshits
states, a population distribution which is narrow in energy
corresponds to a ground state localized on one or very few
well-separated islands, the Lifshits glass. As the temperature
increases the Bose liquid tends to have a rather narrow population
distribution, while the Boltzmann and interacting distributions
quickly spread. Nonetheless, a few important differences may be
noted comparing the two latter cases, as maybe seen in
Figs.~\ref{fig:temperatureAndInteractionEffects}a and
\ref{fig:temperatureAndInteractionEffects}c. While the population
distribution for the Boltzmann case smoothly decreases, with a
long tail, the distribution for the interacting case has a local
minimum at very low $k$, while at larger $k$ goes through a
maximum and then quickly drops to zero. These features can be
understood in the following way. The eigenstates with lowest
energies tend to have higher inverse participations (overlaps),
i.e., the Lifshits states are rather concentrated on a single or
few islands while states at intermediate energies are delocalized
over the whole system. For an interacting system is therefore
preferable to occupy states indexed by states with intermediate
$k$ values, as their larger spatial extension helps to reduce the
interaction energy. At very high energies ($k\gg1$) instead the
states have again very high $O_{kk}$, since they become completely
localized on single sites, and their populations abruptly drop to
zero, as it is too costly energetically to put particles with
repulsive interactions in small regions of space.

\section{Conclusions}\label{sec:conclusions}
In this paper we have considered a bosonic lattice gas in
the presence of Bernoulli disorder, given by randomly-localized identical
scatterers. We have shown that in this case it is possible to
provide precise analytical estimates even for the interacting case.
\revision{We have compared two theoretical schemes,
the simplified multi-orbital Hartree--Fock and the Gross--Pitaevskii
approaches, showing how the first is very accurate in the glassy
regime of strong disorder, but it fails when interactions bring the system into an extended state and the
superfluid fraction reaches values $\rho_{\it sf}\gtrsim 0.1$.}

Further, we have shown analytically that the fractal dimension for
this kind of potential tends to the physical dimension of the
system. As a result, we have introduced a quantity termed
{\it fractional occupation}, which characterizes the typical
extension of the system. When the latter becomes of order $p$,
i.e., the fraction of physical space where scatterers are
absent, the system crosses over from the Lifshits to the Bose
glass. Finally, we have discussed the similarities and differences
between effects due to interaction and temperature. This
question is of fundamental importance for ongoing experiments, since in a
laboratory the two effects are often difficult to distinguish. We
hope that our results will provide new insights in the complex
route towards understanding the interplay between disorder,
interactions, and temperature in quantum mechanical systems.

\ack
\revision{We acknowledge insightful discussions with F. Cuccheti and L.
Sanchez-Palencia. Finantial support from Spanish goverment
Grants  FIS2008-01236, FIS2008-00784 and Consolider Ingenio CDS2006-0019,
Catalan Goverment Grant SGR09-00343 as well as ERC- Advance Grant QUAGATUA and
ERDF (European Regional Develpment Fund) is acknowledged.
M.L also acknowleges Alexander Humbolf Foundation and Hamburg Theory Award.
J.W. and M.B. work was partially supported by the NSF grant DMS 0623941. J.S. is
supported by the Spanish Ministry of Education through the program FPU. J.W.
thanks ICFO and L. Torner for hospitality.}

\appendix
\section{Comparison of the Bernoulli and uniform distributions}
\label{app:convergence} The choice of potential may change the
behaviour and scaling of the ground state energy in a
single-particle system.  We will consider potentials that are
bounded below at $0$ and to simplify notation let $P(x) = P[V_j
\leq x]$ for a given distribution.  Any potential that is bounded
below can be shifted so that the bottom of its support is zero.
Then $P(x) = 0 $ for $x<0$.  For the ground state, we are
interested in the distribution around zero; the behaviour of the
distribution outside of the neighborhood of zero does not affect
limiting behaviour. Assume that the probability of the potential
being less than $\epsilon$ in a neighborhood of zero is given by
$P[V\leq \epsilon] = a + m\epsilon^\gamma$, where $a \in [0,1)$,
$m\geq 0$, and $\gamma > 0$.  Then the expected potential energy
of a ground state supported on the island with each site having
potential less than $\epsilon$ is:

\begin{eqnarray}
E[V|V\leq \epsilon] = \frac{\int_0^\epsilon x\frac{\partial}{\partial
x}(a+mx^\gamma) dx}{a+m\epsilon^\gamma}\nonumber\\
= \frac{m\gamma\epsilon^{\gamma+1}}{2(\gamma +1)(a+m\epsilon\gamma)}.
 \end{eqnarray}

For a specific $\epsilon$ and related probability $P(\epsilon)$, the typical largest island composed by sites where the potential is less than $\epsilon$ has size
\beq
L_{\rm max}\approx\frac{\log \left(P(\epsilon)(1-P(\epsilon))L\right)}{\log P(\epsilon)}
\eeq

Then the energy of the ground state is expected to be
approximately \beq \frac{(|\log
P(\epsilon)|)^{2/D}}{(\log(P(\epsilon)(1-P(\epsilon))L))^{2/D}} +
\frac{m\gamma\epsilon^{\gamma+1}}{2(\gamma +1)(a+m\epsilon\gamma)}
\eeq

For this to be minimized in the large system limit, the order of
each term must be approximately the same, so $\epsilon$ must scale
with $L$ as: \beq \epsilon_L^{\gamma+1} \approx \frac{1}{(\log
L)^{2/D}}. \eeq Plugging this into the kinetic term should give an
approximate rate of convergence \beq\frac{(|\log a +
\frac{m}{(\log L)^{2/D}}|)^{2/D}}{(\log L)^{2/D}} \eeq

For the Bernoulli distribution, $a = p$ and $m=0$, so the rate of
$\log L^{-2/D}$ is recovered.  For the uniform distribution, $a=0$
and $m = \frac{1}{V'}$ and has rate of convergence
$(\frac{\log\log L}{\log L})^{2/D}$ which is slower than the
Bernoulli distribution.  In fact, any distribution with $a \neq
0$, which means the potential can equal exactly zero with positive
probability, must have its associated ground state energy converge
to zero faster than for distributions with $a=0$.  Considering
distributions with $a > 0$ and comparing the cases $m=0$ and
$m>0$, the ground state energy in the former case will be larger
than in the latter case, but asymptotically they will converge to
zero at the same rate. More importantly, in the case where $m=0$,
the longest island is clearly defined whereas its definition
involves an arbitrary (or artificial) cutoff.
in the case where $m > 0$.

\section{Solution of the sMOHF equations}
\label{app:completeSol}
The occupation probabilities $n_k$  may be found by minimizing the total
energy
\beq \label{eq:energy_cross} E=\sum_k n_k \left[E_k+\frac{g}{2}\sum_j
(n_j-\delta_{k,j})\tilde{O}_{k,j}\right], \eeq
where $\tilde{O}=2 O-\mathrm{Diag}(O)$, subject to the constraints:
\beq \sum_k n_k=N,\quad \forall k,\, n_k\geq 0.\eeq
The minimization yields the linear system (one equation for each $k$):
\beq \sum_j n_j
\tilde{O}_{j,k}=\frac{\mu+\gamma_k-E_k}{g}+\frac{\tilde{O}_{k,k}}{2}, \eeq
where Lagrange multipliers $\gamma_k$ and $\mu$ correspond to the
conditions of positivity of each $n_k$ and global normalization,
respectively.
We introduce the vector
\beq
v_k=\left[\frac{\mu-E_k}{g}+\frac{\tilde{O}_{k,k}}{2}\right]_k,
\eeq
which allows us to write the solution as
\beq
\mathbf{n}=\tilde{O}^{-1}\mathbf{v}+\tilde{O}^{-1}\frac{\bgamma}{g}.
\eeq
Now, $\boldsymbol{\gamma}$ and $\mu$ should be chosen such that
the following conditions, named KKT \cite{ConvexOptimization}, are
fulfilled:
\beq \gamma_k\geq 0 \; \& \; \gamma_k n_k=0 \; \& \; \sum_k n_k=N.\eeq
The KKT conditions allow for solving the optimization
problem in the case in which some of the constraints are in the
form of inequalities. These conditions lead to a nonlinear system for
$\bgamma$ and $\mu$.


\section*{References}

\begin{thebibliography}{20}
\bibitem{Kramer93} Kramer B and MacKinnon A 1993 {\it Rep. Prog. Phys.} {\bf
56} 1469.

\bibitem{Anderson1Dexp} Billy J, Josse V, Zuo Z, Bernard A, Hambrecht B, Lugan
P, Cl\'ement D, Sanchez-Palencia L, Bouyer P and Aspect A 2008 {\it Nature} {\bf
453} 891; Roati G, D’Errico C, Fallani L, Fattori M, Fort C, Zaccanti M, Modugno
G, Modugno M and Inguscio M 2008 {\it Nature} {\bf 453} 895.

\bibitem{Anderson3Dexp} Kondov S S, McGehee W R, Zirbel J J and DeMarco B 2011
{\it Science} {\bf 334} 66; Jendrzejewski F, Bernard A, Mueller K, Cheinet
P, Josse V, Piraud M, Pezzé L, Sanchez-Palencia L, Aspect A and Bouyer P 2011
Three-dimensional localization of ultracold atoms in an optical disordered
potential {\it Preprint} cond-mat.other/1108.0137.

\bibitem{Sanchez-Palencia10} Sanchez-Palencia L and Lewenstein M 2010 {\it
Nature Phys.} {\bf 6} 87.

\bibitem{Lewenstein12} Lewenstein M, Sanpera A and Ahufinger V 2012 {\it
Ultracold Atoms in Optical Lattices: Simulating Quantum Many Body Physics}
(Oxford: Oxford University Press) in print.

\bibitem{Giamarchi88} Giamarchi T and Schulz H J 1988 {\it Phys. Rev. B} {\bf
37} 325.

\bibitem{Fisher89} Fisher M P A, Grinstein G and Fisher D S 1989 {\it Phys.
Rev. B} {\bf 40} 546.

\bibitem{Scalettar91} Scalettar R T, Batrouni G G and Zimanyi G T 1991 {\it
Phys. Rev. Lett.} {\bf 66}, 3144.

\bibitem{Damski03} Damski B, Zakrzewski J, Santos L, Zoller P and Lewenstein M
2003 {\it Phys. Rev. Lett.} {\bf 91} 080403.

\bibitem{Lugan07} Lugan P, Cl\'ement D, Bouyer P, Aspect A, Lewenstein M and
Sanchez-Palencia L 2007 {\it Phys. Rev. Lett.} \textbf{98}, 170403.

\bibitem{Lugan07bis} Lugan P, Cl\'ement D, Bouyer P, Aspect A and Sanchez-Palencia
L 2007 {\it Phys. Rev. Lett.} \textbf{99}, 180402.

\bibitem{Altman08} Altman E, Kafri Y, Polkovnikov A and Refael G 2008 {\it Phys.
Rev. Lett.} {\bf 100}, 170402.

\bibitem{Pollet09} Pollet L, Prokof’ev N V, Svistunov B V and Troyer M 2009,
{\it Phys. Rev. Lett.} {\bf 103}, 140402; Gurarie V, Pollet L, Prokof’ev N V,
Svistunov B V and Troyer M 2009 {\it Phys. Rev. B} {\bf 80} 214519.

\bibitem{Gaul11} Gaul C and M\"uller C A 2011 {\it Phys. Rev. A} {\bf 83}
063629.

\bibitem{KirschSurvey} Kirsch W 2007 An invitation to random Schroedinger
operators {\it Preprint} math-ph/0709.3707.

\bibitem{Bishop11} Bishop M and Wehr J 2011 Ground state energy of the
one-dimensional discrete random Schrödinger operator with Bernoulli potential
{\it Preprint} math-ph/1109.4109.

\bibitem{Fallani07} Fallani L, Lye J, Guarrera V, Fort C and Inguscio M 2007
{\it Phys. Rev. Lett.} {\bf 98}, 130404.

\bibitem{Pasienski10} Pasienski M, McKay D, White M and DeMarco B 2010
{\it Nature Phys.} {\bf 6} 677.

\bibitem{Gadway11} Gadway B, Pertot D, Reeves J, Vogt M and Schneble D 2011
{\it Phys. Rev. Lett.} {\bf 107} 145306.

\bibitem{Deissler10} Deissler B, Zaccanti M, Roati G, D'Errico C, Fattori M,
Modugno M, Modugno G and M. Inguscio 2010 {\it Nature Phys.} {\bf 6} 354.

\bibitem{Cederbaum} Cederbaum L S and Streltsov A I 2003 {\it Phys. Lett. A}
{\bf 318} 564;
Cederbaum L S and Streltsov A I 2004 {\it Phys. Rev. A} {\bf
70} 023610;
Streltsov A I and Cederbaum L S 2005 {\it Phys. Rev.
A} {\bf 71} 023612;
Alon O E, Streltsov A I and Cederbaum L S 2007 {\it Phys. Lett. A} {\bf 362}
453;
Alon O E, Streltsov A I and Cederbaum L S 2009 {\it The Multiconfigurational
Time-Dependent Hartree Method for Identical Particles and Mixtures Thereof}
(in {\it Multidimensional Quantum
Dynamics: MCTDH Theory}) ed Meyer H-D, Gatti F and Worth G A
(Eds.), (Weinheim: Wiley-VCH).

\bibitem{Gavish05Massignan06} Gavish U and Castin Y 2005 {\it Phys. Rev. Lett.}
{\bf 95} 020401; Massignan P and Castin Y 2006 {\it Phys. Rev. A} {\bf 74}
013616.

\bibitem{SingleSite} Nelson K, Li X and Weiss D 2007 {\it Nature Phys.} {\bf 3}
556;
Gericke T, W\"urtz P, Reitz D, Langen T and Ott H 2008 {\it Nature Phys.} {\bf
4} 949;
Karski M, F\"orster L, Choi J M, Alt W, Widera A and Meschede D 2009 {\it Phys.
Rev. Lett.} {\bf 102} 053001;
Bakr W S, Gillen J I, Peng A, Foelling S and Greiner M 2009 {\it Nature} {\bf
462} 74.

\bibitem{Antal95} Antal P 1995 {\it Annals of Probability} {\bf 23} 1061.

\bibitem{Simon85} Simon B 1985 {\it J. Stat. Phys.} {\bf 38} 66.

\bibitem{BKT} Berezinskii V L 1972, Sov. Phys. JETP {\bf 34}, 610-616.
Kosterlitz J M and Thouless D J 1973, J. Phys. C {\bf 6}, 1181-1203.

\bibitem{ConvexOptimization} Boyd S and Vandenberghe L 2004 {\it Convex
Optimization} (Cambridge: Cambridge University Press).

\end{thebibliography}
\end{document}